\documentclass[english]{article}
\usepackage[T1]{fontenc}
\usepackage[latin9]{inputenc}
\usepackage[letterpaper]{geometry}
\usepackage{color}
\usepackage{babel}
\usepackage{amsmath}
\usepackage{amsthm}
\usepackage{setspace}

\usepackage[unicode=true,pdfusetitle,
 bookmarks=true,bookmarksnumbered=false,bookmarksopen=false,
 breaklinks=false,pdfborder={0 0 1},backref=false,colorlinks=true]
 {hyperref}

\makeatletter
\newcommand{\lyxaddress}[1]{
	\par {\raggedright #1
	\vspace{1.4em}
	\noindent\par}
}

\usepackage{lmodern}
\usepackage{authblk}

\makeatother

\begin{document}
%%% Comments begin %%%%
\iffalse
::Structure of other essays::

Bold statement in abstract.

One simple idea to present.

Quotes, history.

Well-chosen example to illustrate the general idea.\\

::Content of this essay::

Bold statement: Statements about relativistic effects are misleading

One simple idea to present: It is not obvious what is relativistic,
Einstein's tests are not a test of relativity principle as such.
\pagebreak{}
%%% Comments end %%%%
\fi
%%%%%%%%%%%%%%%%%%%%%%%%%%%%%%%%%%%%%%%
\title{Gravity between Newton and Einstein}

\author[1]{\small Dennis Hansen}
\author[2]{\small Jelle Hartong\footnote{Corresponding author}}
\author[3,4]{\small Niels A. Obers}

\affil[1]{\footnotesize Institut f{\"u}r Theoretische Physik, Eidgen{\"o}ssische Technische Hochschule Z{\"u}rich,

Wolfgang-Pauli-Strasse 27, 8093 Z{\"u}rich, Switzerland.}
\affil[2]{\footnotesize School of Mathematics and Maxwell Institute for Mathematical Sciences,

University of Edinburgh, Peter Guthrie Tait road, Edinburgh EH9 3FD, UK.}

\affil[3]{\footnotesize Nordita, KTH Royal Institute of Technology and Stockholm University, 

Roslagstullsbacken 23, SE-106 91 Stockholm, Sweden.}

\affil[4]{\footnotesize The Niels Bohr Institute, Copenhagen University,

Blegdamsvej 17, DK-2100 Copenhagen \O , Denmark.}
\affil[ ]{\textit{dehansen@phys.ethz.ch, j.hartong@ed.ac.uk, obers@nbi.ku.dk}}

\maketitle
\emph{Keywords}: Gravity, relativity, non-relativity, action, $1/c^2$ expansion.

\begin{abstract}
Statements about relativistic effects are often subtle. In this essay we will demonstrate that the three classical tests of general relativity, namely perihelion precession, deflection of light and gravitational redshift, are passed perfectly by an extension of Newtonian gravity that includes gravitational time dilation effects while retaining a non-relativistic causal structure. This non-relativistic gravity theory arises from a covariant large speed of light expansion of Einstein's theory of gravity that does not assume weak fields and which admits an action principle.

% and since we can incorporate relativistic effects it provides a systematic strong field approximation to general relativity.

%has potential applications to strong gravity effects in astrophysics and cosmology.

% along with a formulation of non-relativistic quantum gravity

\end{abstract}

\lyxaddress{\begin{center}
\emph{Essay written for the Gravity Research Foundation 2019\\Awards for Essays on Gravitation.}
\par\end{center}}
%
%\subsection*{Tests of GR}
\newpage
In 1916 Albert Einstein formulated the three classical tests of his General Relativity, namely \cite{Einstein1916}:
\begin{enumerate}
\item the precession of the perihelion of Mercury's orbit
\item the deflection of light by the Sun, and 
\item the gravitational redshift effect.
\end{enumerate}
Subsequently all of these predictions have been tested experimentally with wonderful agreement.
Originally these tests were devised as it was known that Newton's law of universal gravitation gave inaccurate predictions.
All three tests are described in the general relativity framework by studying the Schwarzschild solution and its geodesics.
%The standard narrative is that the origin of these three effects is relativistic in nature, 
%as is naively suggested by the explicit appearance of $1/c^2$ in the corresponding general relativity results.
%that forms an extension of Newtonian gravity by including
%
%computing these effects does not require Einstein's notion of special relativity:
% and that they can be fully accounted for using a theory of non-relativistic gravity
%
%
We will argue that these effects do not require Einstein's notion of special relativity. The three classical tests are actually solely \emph{strong gravitational field effects}.
This follows from an extension of Newtonian gravitation that includes time dilation \cite{Hansen:2018ofj}.
This novel theory is obtained from a covariant off shell $1/c^{2}$ expansion of general relativity,
leading naturally  to a gravity theory with manifest non-relativistic symmetries \cite{VandenBleeken:2017rij, VandenBleeken:2019gqa}. 
%DH: 30/03 I changed the sentence here to shorten it, but it also changes the statement.
It is much richer than Newtonian gravity and can account for many results we know from general relativity.
This opens the way for a more nuanced discussion about the status of gravity and its quantum version in the non-relativistic domain.

The Einstein equivalence principle says that locally the laws of physics reduce to those of special relativity. On the other hand, insisting on local Galilean relativity will lead to Newton--Cartan geometry, pioneered by Cartan \cite{Cartan1,Cartan2} and further developed in e.g. \cite{Eisenhart, Trautman63}.
The Galilean algebra consists of time and space translations $H,\,P_a$, Galilean boosts $G_a$ and rotations $J_{ab}=-J_{ba}$, where $a,b=1,\ldots,d$ are spatial indices.
The relevant local symmetry algebra is actually its central extension, known as the Bargmann algebra, where the central charge $N$ (appearing in $[P_a,G_b]=N\delta_{ab}$) can be interpreted as mass.
One can obtain Newton--Cartan geometry by gauging the Bargmann algebra \cite{Andringa:2010it,Hartong:2015zia},
which in turn can be obtained via the In\"on\"u--Wigner contraction of the direct sum of the Poincar\'e algebra and a $U(1)$ generator. Since not all relativistic systems possess a $U(1)$ symmetry
this has the serious drawback that it is not a generic limit that always exists.

%\footnote{Since the algebra strictly lives in the tangent space, what is meant by $1/c$ here is a dimensionless number proportional to the slope of the light cone.\label{footnote}}

Another way to obtain the non-relativistic limit of relativistic physics is to re-instate factors of $c$ and to perform a $1/c^2$ expansion up to the desired order. This approach can always be done and is thus generic, but as we will see it is not necessarily equivalent to the above mentioned Bargmann limit. The Poincar\'e algebra has generators $T_A=\{H,P_a,B_a,J_{ab}\}$, where $B_a$ are the Lorentz boosts.
We first re-instate all factors of $c$ explicitly in the commutation relations and 
tensor the generators with the polynomial ring in $\sigma \equiv 1/c^2$.
We then obtain the basis of generators $T^{(n)}_A\equiv T_A \otimes \sigma^n$, where $n\ge 0$ is the level.
This is a graded algebra with nonzero commutation relations of the form 
\begin{equation}\label{eq:commutation_relations_ring}
\left[H^{(m)},B^{(n)}_a\right]=P^{(m+n)}_a\,,\quad\left[P^{(m)}_a,B^{(n)}_b\right]=\delta_{ab}H^{(m+n+1)}\,,\quad\left[B^{(m)}_a,B^{(n)}_b\right]=-J^{(m+n+1)}_{ab}\,,
\end{equation}
where we left out the commutators with $J_{ab}^{(m)}$. We can quotient this algebra by setting to zero all generators with level $n>L$ for some $L$ \cite{Khasanov:2011jr}.
At level $n=0$ the algebra is isomorphic to the Galilean algebra.
Including the level $n=1$ generators $\{H^{(1)},P_a^{(1)},B_a^{(1)},J_{ab}^{(1)}\}$ one finds a commutator $[P_a,G_b]=H^{(1)}\delta_{ab}$, except that now $[H^{(1)},G_a]=P^{(1)}_a$. Hence $H^{(1)}$ is not central like in the Bargmann algebra, and thus the latter is {\it not} a subalgebra.
This has severe implications: the result of gauging the Bargmann algebra will in general \emph{not coincide}
with the $1/c^2$ expansion of general relativity \cite{Hansen:2018ofj}!

To see how strong field effects are encoded into the non-relativistic limit of general relativity,
we define the following covariant $1/c^2$ expansion (where $c$ is the slope of the light cone in tangent space) of the Lorentzian metric\footnote{This follows from writing $g_{\mu\nu}=-c^2E^0_\mu E^0_\mu+\delta_{ab}E^a_\mu E^b_\nu$ and Taylor expanding the vielbeins $E^0_\mu$ and $E^a_\mu$ in $\sigma=c^{-2}$.} $g_{\mu\nu}$ \cite{VandenBleeken:2017rij,Dautcourt}:
\begin{eqnarray}
g_{\mu\nu} & = &-c^{2}\tau_{\mu}\tau_{\nu}+\overline{h}_{\mu\nu}-\frac{1}{c^{2}}\Phi_{\mu\nu}+\cdots\,,  \\
g^{\mu\nu} & = & h^{\mu\nu}-\frac{1}{c^{2}}\hat{v}^{\mu}\hat{v}^{\nu}+\frac{1}{c^{2}}h^{\mu\rho}h^{\nu\sigma}\Phi_{\rho\sigma}+\frac{1}{c^{4}}2\hat{v}^{\mu}\hat{v}^{\nu}\tilde{\Phi}+\cdots\,,
\end{eqnarray}
where $\tilde \Phi=-\hat v^\mu\hat v^\nu\bar h_{\mu\nu}/2$. The omitted terms will not contribute to our calculations. Covariance requires diffeomorphisms that are analytic in $1/c^2$. Expanding Einstein's equations in $1/c^2$ tells us that at leading order $\tau\wedge d\tau=0$, so that the clock 1-form $\tau=\tau_{\mu}dx^\mu$ describes a foliation of spacetime. 
When $d\tau\neq 0$ two observers with worldlines $\gamma_{1},\,\gamma_{2}$ need not experience
the same lapse of time, $\int_{\gamma_{1}}\tau\neq\int_{\gamma_{2}}\tau$.
In Galilean relativity space is measured by the inverse spatial metric $h^{\mu\nu}$
with signature $(0,1,\ldots,1)$ satisfying $h^{\mu\nu}\tau_{\nu}=0$.
The tensors $\hat{v}^\mu \equiv v^\mu - h^{\mu \nu} m_\nu $ and 
$\bar{h}_{\mu\nu} \equiv h_{\mu \nu} - m_\mu \tau_\nu - m_\nu \tau_\mu$ 
are projective inverses and furthermore involve the field $m_\mu$,
which is important as $m_0$ is the Newtonian potential.
They satisfy the completeness relations 
\begin{equation}\label{eq:completeness_relations}
-\hat{v}^\mu\tau_\nu+h^{\mu\lambda}\bar{h}_{\lambda\nu}=\delta^\mu_\nu
\qquad \textrm{and}\qquad \hat{v}^\mu\tau_\mu=-1\,.
\end{equation}
The geometry is described by $\tau_\mu$ and $h^{\mu\nu}$. The subleading fields $m_\mu$ and $h^{\mu\rho}h^{\nu\sigma}\Phi_{\rho\sigma}$ are gauge fields on the geometry.

A natural choice for the affine connection is \cite{Hartong:2015zia}
\begin{equation}\label{eq:nc_connection}
\bar{\Gamma}_{\mu\nu}^{\lambda}=-\hat{v}^{\lambda}\partial_{\mu}\tau_{\nu}
+\frac{1}{2}h^{\lambda\sigma}\left(\partial_{\mu}\overline{h}_{\nu\sigma}+\partial_{\nu}\overline{h}_{\mu\sigma}-\partial_{\sigma}\overline{h}_{\mu\nu}\right)\,.
\end{equation}
This connection is Newton--Cartan metric compatible, i.e. $\bar{\nabla}_{\mu}\tau_{\nu}=\bar{\nabla}_{\mu}h^{\nu\rho}=0$ where $\bar\nabla$ is the covariant derivative associated to \eqref{eq:nc_connection}.
%The condition $\bar{\nabla}_{\mu}\tau_{\nu}=0$  implies that
Notice that the connection has torsion $\bar{\Gamma}_{[\mu\nu]}^{\lambda}=-\hat{v}^{\lambda}\partial_{[\mu}\tau_{\nu]}$ when $d\tau\neq 0$.
Our hand is thus forced: when repackaging gravity in the geometric language of Newton--Cartan geometry, torsion arises naturally. The most general solution to $\tau\wedge d\tau=0$ is $\tau=N\mathrm{d}T$, where $T$ is a time function and $N$ is the non-relativistic lapse function.
As in general relativity we can always use diffeomorphisms to choose $T=t$ as the time coordinate.
However, in Newton-Cartan geometry there is not enough freedom to set $N$ to unity:
\emph{the lapse function is physical}. Setting $d\tau=0$ corresponds to a weak gravitational field.

We are now in a position to expand the Einstein--Hilbert action in powers of $1/c^2$. Since we expand the metric up to subleading orders we will do the same for the action. Schematically the Einstein--Hilbert Lagrangian expands as $\mathcal{L}_{\text{EH}}=c^4\mathcal{L}_{\text{LO}}+c^2\mathcal{L}_{\text{NLO}}+\cdots$ \cite{Hansen:2019svu}. It can be shown that the equations of motion of the leading order Lagrangian $\mathcal{L}_{\text{LO}}$ are identical to the equations of motion of the subleading fields appearing in the next-to-leading order Lagrangian $\mathcal{L}_{\text{NLO}}$. We can thus throw away $\mathcal{L}_{\text{LO}}$ and work with $\mathcal{L}_{\text{NLO}}$. This defines the theory of non-relativistic gravity\footnote{Continuing the $1/c^2$ expansion to even higher orders will bring in relativistic effects to the theory,
but we shall refrain from studying them in this essay.} and the action is  \cite{Hansen:2018ofj}:
\begin{multline}
S = -\frac{1}{16\pi G}\int \mathrm{d}^{d+1}xe\Big [\hat v^\mu\hat v^\nu\bar R_{\mu\nu}-\tilde\Phi h^{\mu\nu}\bar R_{\mu\nu}
-\Phi_{\mu\nu}h^{\mu\rho}h^{\nu\sigma}\left(\bar R_{\rho\sigma}-a_\rho a_\sigma-\bar\nabla_{\rho}a_{\sigma}\right)\\
\qquad\qquad\qquad\qquad\quad+\frac{1}{2}\Phi_{\mu\nu}h^{\mu\nu}\left[h^{\rho\sigma}\bar R_{\rho\sigma}-2h^{\rho\sigma}\left(a_\rho a_\sigma+\bar\nabla_{\rho}a_{\sigma}\right)\right]\Big]
+S_{\mathrm{M}}\,, \label{eq:typeII_action}
\end{multline}
where $e=\left(-\mathrm{det}(-\tau_\mu\tau_\nu+h_{\mu\nu})\right)^{1/2}$, $a_\mu\equiv2\hat{v}^{\rho}\partial_{[\rho}\tau_{\mu]}$, $\bar R_{\mu\nu}$ the Ricci tensor associated to $\bar{\Gamma}_{\mu\nu}^{\lambda}$ and $S_{\mathrm{M}}$ contains possible matter couplings. 

In the absence of matter sources for $\tau_\mu$ and demanding that $\tau_\mu$ is nowhere vanishing the equations of motion force $d\tau=0$. In this case the field $\Phi_{\mu\nu}$ decouples from the equations of motion for $m_\mu$ and $h_{\mu\nu}$. An example of a simple matter Lagrangian that does not source $\tau_\mu$ is $\mathcal{L}_{\mathrm{M}}=-(d-2)e\rho/2$ where $\rho$ is a mass density. The equations of motion lead to 
\begin{equation}\label{eq:Poisson_geometric}
\bar{R}_{\mu\nu}=8\pi G\frac{d-2}{d-1}\rho\tau_{\mu}\tau_{\nu}\,,
\end{equation}
with $d\tau=0$ and $\Phi_{\mu\nu}$ appearing in a decoupled equation. This is the geometrized diffeomorphism covariant version of the Poisson equation \cite{Trautman63}. 
The theory however allows for more general geometries in which $d\tau\neq 0$. These are the non-relativistic gravity solutions that encode strong field effects. We will now focus on those, i.e. we will assume that there is a matter distribution that sources $d\tau\neq 0$ (such as a spherical non-rotating fluid star) outside of which we have a vacuum solution describing time dilation effects \cite{Hansen:2020pqs}.
%DH: 30/03 Must be self-contained, also to save space here.
%The existence of such matter configurations in the non-relativistic setting has been worked out in \cite{Hansen:2020pqs}.

To study the motion of massive particles on a background with $d\tau\neq 0$ we must perform an expansion of the relativistic proper time worldline action.
We need to expand the embedding scalars as $\tau_\mu\dot X^\mu=\frac{1}{c}\tau_\mu\dot y^\mu+\cdots$ and $h_{\mu\nu}\dot X^\mu=h_{\mu\nu}\dot x^\mu+\cdots$. The resulting reparameterization invariant action is 
\begin{equation}\label{eq:worldline_action_typeII}
S \propto\int d\lambda\left[\left(\frac{\mathrm{d}}{\mathrm{d}\lambda}\left(\tau_\mu y^\mu\right)- \tau_\mu y^\mu a_\nu \dot x^\nu \right)^2-{h}_{\mu\nu}\dot x^\mu\dot x^\nu\right]^{1/2}\,.
\end{equation}
In a gauge in which the term in square brackets in the integrand of the action \eqref{eq:worldline_action_typeII} is a constant ($-2\mathcal{E}>0$) on shell
and using $\tau_\mu=N\partial_\mu T$, one finds the equations of motions
\begin{eqnarray}
\frac{\mathrm{d}}{\mathrm{d}\lambda}\left(\tau_\mu y^\mu\right)- \tau_\mu y^\mu a_\nu \dot x^\nu&=&N^2\frac{\mathrm{d}}{\mathrm{d}\lambda}\left(y^\mu\partial_\mu T\right)=1\,,\\
\ddot{x}^{\lambda}+\bar{\Gamma}_{\mu\nu}^{\lambda}\dot{x}^{\mu}\dot{x}^{\nu}&=&\frac{1}{2}h^{\lambda\sigma}\partial_{\sigma}N^{-2}\,,\label{eq:EOM_geodesic1}\\
\mathcal{E}&=&-\frac{1}{2}\frac{1}{N^2}+\frac{1}{2}h_{\mu\nu}\dot x^\mu\dot x^\nu\,.\label{eq:EOM_geodesic2}
\end{eqnarray}
The last term in \eqref{eq:EOM_geodesic1} is the negative gradient of the potential energy $-N^{-2}/2$. The term
$\tau_\mu y^\mu$ acts as a Lagrange multiplier implementing conservation of energy $\mathcal{E}$ in \eqref{eq:EOM_geodesic2}. We thus see that the expansion of timelike geodesics leads to bound ($\mathcal{E}<0$) non-relativistic particles with potential energy $-N^{-2}/2$. The $1/c^2$ expansion for null geodesics leads to identical equations except with $\mathcal{E}=0$.

It can be shown that the most general 4-dimensional static spherically symmetric vacuum solution to
the equations of motion of \eqref{eq:typeII_action} is given by 
\begin{eqnarray}\label{TNC_4D_spherical_symmetric_solution}
\tau_{\mu} & = & \sqrt{1-\frac{r_s}{r}}\delta_{\mu}^{0}\,,\\
h^{\mu\nu} & = & \mathrm{diag}\left(0,\left(1-\frac{r_s}{r}\right),1/r^2,1/\left(r^2\sin^2\theta\right)\right)\,,
\end{eqnarray}
with $m_\mu=0$ and $\Phi_{\mu\nu}=0$ and $r_s$ an integration constant \cite{Hansen:2020pqs} where the coordinates are $x^\mu=(t,r,\theta,\phi)$. We can also obtain the above geometry by performing a $1/c^2$ expansion of the Schwarzschild solution, treating the Schwarzschild radius $r_s$ as a $c$-independent integration constant \cite{VandenBleeken:2017rij,VandenBleeken:2019gqa}. This expansion terminates after one order, so that it is a  solution of non-relativistic gravity. The time dilation is described by the lapse function $N(r)=\sqrt{1-{r_s}/{r}}$, which is exactly the same as in general relativity. This proves that our non-relativistic gravity theory passes the third classical test as it is a pure time-dilation effect. For motion in the equatorial plane ($\theta=\pi/2$) equations \eqref{eq:EOM_geodesic1} and \eqref{eq:EOM_geodesic2} lead to 
\begin{equation}\label{eq:geodesic_schwarzschild_effective}
\left(\frac{\mathrm{d}r}{\mathrm{d}\phi}\right)^2=\frac{r^4}{b^2}-\left(1-\frac{r_s}{r}\right)\left(\kappa\frac{r^4}{a^2}+r^2\right)\,,
\end{equation}
where we treat the radial coordinate $r(\phi)$ as a function of the angular variable $\phi$. The parameters $a$ and $b$ have dimension of length and depend on the angular momentum and energy of the particle.
This equation is equivalent to the geodesic equation for motion in the equatorial plane for both massive ($\kappa=1$) and massless ($\kappa=0$) particles in general relativity \cite{Carroll:2004st}. We have thus demonstrated that the first and second classical tests are also passed by the theory described by the action \eqref{eq:typeII_action}!

%where $r_s=2GM/c^2$ is the Schwarzschild radius with $G$ Newton's constant, $M$ the mass of the sun, $c$ the speed of light. 

%Solutions of the equations of motion of \eqref{eq:typeII_action} are exact solutions of general relativity provided the latter admit a 1/c^2$ expansion that terminates at leading or subleading order  \cite{VandenBleeken:2019gqa}.

In general solutions of the equations of motion of \eqref{eq:typeII_action} are exact/approximate solutions of general relativity if the latter's $1/c^2$ expansion terminates/extends beyond subleading order \cite{VandenBleeken:2019gqa}. For example FRW-type spacetimes and the associated Friedmann equations are fully described by non-relativistic gravity coupled to non-relativistic fluids \cite{Hansen:2020pqs}. However the nature of the $1/c^2$ expansion is such that the equations of motion of \eqref{eq:typeII_action} do not admit gravitational wave solutions:
they are true relativistic phenomena. 
%Adding gravitational waves to his list is indeed what Einstein should have done already in 1916.
%DH 29/03: I think we should leave this out, apparantly Einstein was not convinced of their existence until much later. So a complicated discussion.

Finally, the $1/c^2$ expansion of general relativity \emph{is} a simplification even if we have many more fields.
This is because there is a preferred foliation of time and interactions become instantaneous.
For the same reason the quantum version of non-relativistic gravity might be more attainable than its notoriously difficult relativistic parent. In view of the Bronstein cube of $Gc\hbar$-physics, where one typically approaches  (relativistic) quantum gravity
from the quantum field theory or general relativity corner, this could open up a third road by approaching it
from the non-relativistic quantum gravity corner.

\bibliographystyle{unsrt}
\bibliography{Lifshitzhydro_g_DH}

\end{document}